\font \sevenrm=cmr7
\font \fiverm=cmr5
\documentclass[12pt]{amsart}
\input epsf.sty
\usepackage[frenchb]{babel}
\usepackage{amsfonts}
\usepackage{amssymb}
\usepackage{amsmath}
\usepackage{amsthm}
\usepackage{amscd}
\usepackage[all]{xy} 
\usepackage{epsfig,exscale}
\usepackage{color}
\usepackage{graphicx}
\usepackage{xspace}
\usepackage{axodraw}
\usepackage{colordvi}

\newcommand{\nc}{\newcommand}

{\everymath{\displaystyle\everymath{}}\array}%
{\endarray}
\newtheorem{theorem}{Th\'eor\`eme}
\newtheorem{definition}{D\'efinition}
\newtheorem{corollary}{Corollaire}

\newtheorem{proposition}{Proposition}

\newtheorem{remark}{Remarque}
\nc{\comment}[1]{[[{\tt #1}]] }
\nc{\Cal}[1]{{\mathcal {#1}}}
\nc{\mop}[1]{\mathop{\hbox {\rm #1} }\nolimits}
\nc{\gmop}[1]{\mathop{\hbox {\bf #1} }\nolimits}
\def\starz{{\displaystyle\mathop{\star}\limits}_z}
\nc{\smop}[1]{\mathop{\hbox {\sevenrm #1} }\nolimits}
\nc{\ssmop}[1]{\mathop{\hbox {\fiverm #1} }\nolimits}
\nc{\mopl}[1]{\mathop{\hbox {\rm #1} }\limits}
\nc{\smopl}[1]{\mathop{\hbox {\sevenrm #1} }\limits}
\nc{\ssmopl}[1]{\mathop{\hbox {\fiverm #1} }\limits}
\nc{\frakg}{{\frak g}}
\nc{\g}[1]{{\frak {#1}}}
\def \restr#1{\mathstrut_{\textstyle |}\raise-6pt\hbox{$\scriptstyle #1$}}
\def \srestr#1{\mathstrut_{\scriptstyle |}\hbox to
-1.5pt{}\raise-4pt\hbox{$\scriptscriptstyle #1$}}
\nc{\wt}{\widetilde} \nc{\wh}{\widehat}
\nc{\redtext}[1]{\textcolor{red}{#1}}
\nc{\bluetext}[1]{\textcolor{blue}{#1}}
\nc\fleche[1]{\mathop{\hbox to #1 mm{\rightarrowfill}}\limits}
\nc{\ignore}[1]{}
\def\semi{\mathrel{\times}\kern -.85pt\joinrel\mathrel{\raise
1.4pt\hbox{${\scriptscriptstyle |}$}}}
\nc\R{{\mathbb R}}
\nc\N{{\mathbb N}}
\nc\inver{^{-1}}
\nc\point{\hbox{\bf .}}
\nc\un{\hbox{\bf 1}}
\def\racine{{\scalebox{0.3}{ 
\begin{picture}(12,12)(38,-38)
\SetWidth{0.5} \SetColor{Black} \Vertex(45,-33){5.66}
\end{picture}}}}

\def\diagramme #1{\vskip 4mm \centerline {#1} \vskip 4mm}
\begin{document}
\title{
{Groupes de renormalisation pour deux alg\`ebres de Hopf en produit semi-direct}}
\author{Mohamed Belhaj Mohamed}
\address{{Universit\'e Blaise Pascal,
         laboratoire de math\'ematiques UMR 6620,
         63177 Aubi\`ere, France}\\
          {laboratoire de math\'ematiques physique fonctions sp\'eciales et applications, universit\'e de sousse, rue Lamine Abassi 4011 H. Sousse,  Tunisie}}      
         \email{Mohamed.Belhaj@math.univ-bpclermont.fr}
\date{ 05 juillet 2012}
\noindent{\footnotesize{${}\phantom{a}$ }}
\begin{abstract}
Nous consid\'erons deux alg\`ebres de Hopf gradu\'ees connexes en interaction, l'une \'etant un comodule-cog\`ebre sur l'autre. Nous montrons comment d\'efinir l'analogue du groupe de renormalisation et de la fonction B\^eta de Connes-Kreimer lorsque la bid\'erivation de graduation est remplac\'ee par une bid\'erivation provenant d'un caract\`ere infinit\'esimal de la deuxi\`eme alg\`ebre de Hopf.\\

\noindent
{\sc Abstract:} 
We consider two interacting connected graded Hopf algebras, the former being a comodule-coalgebra on the latter. We show how to define analogues of Connes-Kreimer's renormalization group and Beta function, when the graduation operator is replaced by any biderivation coming from an infinitesimal character of the second Hopf algebra.  
\end{abstract}
\maketitle
\tableofcontents
\section{Introduction}
D. Kreimer a montr\'e \`a la fin des ann\'ees 90 (\cite{DK98}) que les graphes de Feynman en th\'eorie quantique des champs s'organisent en une alg\`ebre de Hopf gradu\'ee connexe. Sur toute alg\`ebre de Hopf $\Cal H$ de ce type il est possible  de d\'ecrire un proc\'ed\'e de renormalisation directement apparent\'e \`a l'algorithme de Bogoliubov, Parasiuk, Hepp et Zimmermann (BPHZ) (\cite{nbsp}, \cite{wz}). Le cadre est le suivant : pour toute alg\`ebre commutative unitaire $\Cal A$ munie d'un \emph{sch\'ema de renormalisation}, c'est-\`a-dire d'une d\'ecomposition $\Cal A = \Cal A_- \oplus \Cal A_+$ o\`u $\Cal A_-$ et $\Cal A_+$ sont deux sous-alg\`ebres, avec l'unit\'e dans $\Cal A_+$, tout caract\`ere $ \varphi : \Cal H \rightarrow \Cal A$ admet une unique \emph{d\'ecomposition de Birkhoff} en deux caract\`eres $\varphi_-$ et $\varphi_+$ :
$$\varphi = \varphi_-^{\ast -1} \ast \varphi_+$$
o\`u $\varphi_-(\mop {Ker} \varepsilon) \subset \Cal A_-$ et $\varphi_+(\Cal H) \subset \Cal A_+$ (on d\'esigne par $\varepsilon$ la counit\'e). L'etoile $\ast$ d\'esigne le produit de convolution. Un exemple de sch\'ema de renormalisation est donn\'e par les series de Laurent en une varianble complexe $z$, $\Cal A_+$ d\'esigne alors $\mathbb{C}[[z]]$ et $\Cal A_-$ d\'esigne l'espace $z^{-1}\mathbb{C}[z^{-1}]$ des polynomes en $z^{-1}$sans terme constant (\emph{sch\'ema minimal}). La valeur renormalis\'ee du caract\`ere $\varphi$ est d\'efinie par $\varphi_+(0)$, qui par d\'efinition existe. L'operateur de graduation $Y : \Cal H \rightarrow \Cal H$, donn\'e par $Y(x) = nx$ pour $x$ homog\`ene de degr\'e $n$, est une bid\'erivation de $\Cal H$. On en d\'eduit une action de $\mathbb{C}$ sur le groupe $ G_\Cal A$ des caract\`eres de $\Cal H$ dans $\Cal A = \mathbb{C}[z^{-1},z]]$, donn\'ee par :
$$\varphi_t(x)(z)= e^{tz\left| x \right|} \varphi(x)(z).$$
L'ensemble des caract\`eres locaux est d\'efini par :
$$ G_\Cal A ^{loc} = \{ \varphi \in G_{\Cal A}\;\; \text{tel que : }\;\; \frac{d}{dt}(\varphi_t) =0 \}.$$
Le groupe de renormalisation d'un caract\`ere local $\varphi$ \cite{ad01} est d\'efini par : 
\begin{eqnarray*}
F_{t}(\varphi) (x)= \lim_{z\longrightarrow0} (\varphi^{\ast-1} \ast \varphi_t) (x) (z).
\end{eqnarray*}
La fonction B\^eta est le g\'en\'erateur de ce groupe \`a un param\`etre :
$$\beta (\varphi)(x) := \frac{d}{dt}_{ | t=0} F_{t}(\varphi)(x).$$
L'objectif de ce travail est de d\'efinir des objets analogues pour d'autres bid\'erivations que la graduation $Y$. Une famille de bid\'erivations appara\^\i t dans la situation suivante : On suppose qu'il existe une deuxi\`eme alg\`ebre de Hopf gradu\'ee connexe $\Cal K$ interagissant avec $\Cal H$. Plus pr\'ecis\'ement on suppose qu'il existe une coaction $\Phi:\Cal H \longrightarrow \Cal K \otimes\Cal H $  qui est en m\^eme temps un morphisme d'alg\`ebres gradu\'ees, et telle que :
\begin{equation}
(\mop{Id}_{\Cal K}\otimes\Delta_{\Cal H})\circ\Phi=m^{1,3}\circ(\Phi\otimes\Phi)\circ\Delta_{\Cal H},
\end{equation}
o\`u\;\; $m^{1,3}:\Cal K \otimes\Cal H\otimes \Cal K \otimes \Cal H
\longrightarrow \Cal K\otimes\Cal H\otimes\Cal H$ \;\; est d\'efini par :
$$
m^{1,3}(a\otimes b\otimes c\otimes d)=ac\otimes b\otimes d,
$$
et $ \Phi $ s'exprime en notation de Sweedler pour tout $ x \in \Cal H$ par :
\begin{equation}
\Phi (x) = \sum_{(x)} x_0 \otimes x_1 = \un_\Cal K \otimes x + \sum_{(x)} x^{(')} \otimes x^{('')}.
\end{equation} 
Cette situation se rencontre naturellement dans le cas de l'alg\`ebre de Hopf $\Cal H_{CK}$ des arbres enracin\'es \cite{ckm}, et dans le cas plus g\'en\'eral de l'alg\`ebre de Hopf des graphes de Feynman orient\'es sans cycles \cite{Dm11}. Le groupe $G_\Cal A ^{\Cal K}$ des caract\`eres de $\Cal K$ (\`a valeurs dans $\Cal A$) agit alors par automorphismes sur le groupe $G_\Cal A $ des caract\`eres de $\Cal H$. Tout caract\`ere infinit\'esimal $\alpha : \Cal K \rightarrow \Cal A$ d\'efinit alors une bid\'erivation $B_\alpha$ de l'alg\`ebre de Hopf $\Cal H$, qui peut jouer le r\^ole de la graduation $Y$.\\
Nous montrons que les caract\`eres locaux, le groupe de renormalisation et le fonction B\^eta peuvent \^etre d\'efinis de la m\^eme mani\`ere que pour la bid\'erivation $Y$. Soit $S$ l'antipode de l'alg\`ebre de Hopf $\Cal H$. L'analogue $\varphi \longmapsto \varphi \circ E_\alpha$ de la composition \`a droite par l'operateur de Dynkin $S \ast Y$ n'est toutefois pas une bijection des caract\`eres de $\Cal H$ vers les caract\`eres infinitis\'emaux. Cela vient du fait que $\mop{Ker} B_\alpha$ est non trivial (il contient tous les \'el\'ements primitifs), contrairement \`a $\mop{Ker} Y$ qui se r\'eduit \`a l'unit\'e de $\Cal H$.\\

\noindent
{\bf Remerciements :} Je remercie vivement mes directeurs de th\`ese Mrs Dominique Manchon et Mohamed Selmi. Le pr\'esent travail b\'en\'eficie du soutien du projet CMCU Utique Num\'ero 12G1502. 
\section{Rappels sur les alg\`ebres de Hopf et la renormalisation}
\subsection {Alg\`ebres, cog\`ebres et big\`ebres }
Dans toute la suite, on d\'esigne par $k$ un corps. 
\begin{definition} Une $k$-alg\`ebre unitaire est un triplet $(\Cal A; m; u) $ o\`u $\Cal A$ est un $k$-espace vectoriel et
$$\ m :\Cal A \otimes \Cal A \longrightarrow \Cal A  , \hskip 6mm u : k\longrightarrow \Cal A ,$$
sont deux applications lin\'eaires satisfaisant les deux axiomes suivants :
\begin{enumerate}
\item Associativit\'e :
$$m\circ(m \otimes Id) = m\circ( Id \otimes m).$$
\item Unit\'e :$$m\circ(u \otimes Id) = Id = m\circ( Id \otimes u).$$
\end{enumerate}
\end{definition}
\begin{definition}
Une cog\`ebre co-unitaire est un triplet $(\Cal C ; \Delta ;\varepsilon)$ o\`u $\Cal C$ est un $k$-espace vectoriel et $\Delta :\Cal C \longrightarrow \Cal C \otimes \Cal C$ (coproduit), $\varepsilon : \Cal C \longrightarrow k$ (counit\'e) sont deux applications lin\'eaires satisfaisant les deux axiomes suivants :
\begin{enumerate}
\item Coassociativit\'e :
$$ (\Delta \otimes Id) \circ \Delta = (Id \otimes \Delta) \circ \Delta.$$
\item Counit\'e :$$(\varepsilon \otimes Id) \circ \Delta = Id_{\Cal C} = (Id \otimes \varepsilon) \circ \Delta.$$
\end{enumerate}
\end{definition}
\textbf{Notation de Sweedler}. Le coproduit d'un \'el\'ement est donc une somme finie d'\'el\'ements ind\'ecomposables. Pour d\'ecrire le coproduit, on utilise la notation suivante :
\begin{equation}
\Delta (x) = \sum_{(x)} x_{1} \otimes x_{2}.
\end{equation}
\begin{definition}
Une big\`ebre est une famille $(\Cal H , m ,u ,\Delta , \varepsilon)$ telle que :
\begin{enumerate}
\item $(\Cal H , m ,u )$ est une alg\`ebre unitaire.
\item $(\Cal H , \Delta , \varepsilon)$ est une cog\`ebre co-unitaire.
\item $ \Delta$ et $\varepsilon$ sont des morphismes d'alg\`ebres unitaires ou, de mani\`ere \'equivalente $m$ et $u$ sont des morphismes de cog\`ebres co-unitaires. 
\end{enumerate} 
\end{definition}
\subsection {Dual gradu\'e}
Soit $V = \bigoplus_{ n \geq 0} V_n$ un espace gradu\'e. Soit $n \geq 0$. Alors $V^\ast_n$ s'identifie au sous espace suivant de $V^\ast$ :
$$ V^\ast_n \approx \{ f \in V^\ast / f (V_k)(0)\;\; si\; k \neq n\}.$$
Par la suite, on identifiera les deux et on pourra \'ecrire $ V^\ast_n \subseteq V^\ast $.
\begin{definition}
Soit $V$ un espace gradu\'e. Le dual gradu\'e de $V$ est le sous-espace suivant de $V^\ast$ :
$$V^\circ := \bigoplus^\infty_{ n = 0} V^\ast_n .$$
\end{definition}
\begin{remark}
Lorsque chaque $ V_n$ est de dimension finie, $V^{\circ \circ}$ est isomorphe \`a $V$ comme espace gradu\'e.
\end{remark}
\subsection {Convolution et alg\`ebres de Hopf}
\begin{proposition}
Soit $ \Cal C = (\Cal C ; \Delta ;\varepsilon)$ une cog\`ebre co-unitaire et $\Cal A = (\Cal A ; m ;u)$ une alg\`ebre unitaire. L'espace vectoriel $ \mop{Hom}(\Cal C ,\Cal A)$ est munie d'une structure d'alg\`ebre de la mani\`ere suivante : si $f , g \in \mop{Hom}(\Cal C ,\Cal A)$, $$ f \ast g = m \circ ( f \otimes g) \circ \Delta.$$
Autrement dit, pour tout $x \in \Cal C$ : $$ f \ast g (x) = \sum_{(x)} f (x_{1}) g (x_{2}).$$
Ce produit est appel\'e produit de convolution. L'unit\'e est l'application $ i: x \longrightarrow \varepsilon (x) \un_{\Cal A}.$
\end{proposition}
\begin{definition}
Soit $\Cal H $ une big\`ebre. On dira que $\Cal H $ est une alg\`ebre de Hopf si $ Id_{\Cal H}$ poss\`ede un inverse dans l'alg\`ebre de convolution $\mop{Hom}(\Cal H ,\Cal H)$. L'unique inverse de $Id_{\Cal H}$ est appel\'e antipode de $\Cal H$ et il est not\'e en g\'en\'eral $ S$. Autrement dit, $\Cal H$ est une alg\`ebre de Hopf s'il existe une application lin\'eaire $S: \Cal H \longrightarrow \Cal H$ telle que le diagramme suivant commute :
\diagramme{
\xymatrix{
& \Cal H\otimes \Cal H \ar[rr]^{S \otimes Id_{\Cal H}}&& \Cal H\otimes \Cal H \ar[dr]^{m}&\\
\Cal H\ar[ur]^{\Delta} \ar[rr]^{\varepsilon}\ar[dr]^{\Delta}
&& k \ar[rr]^{u} && \Cal H\\
& \Cal H\otimes \Cal H \ar[rr]^{ Id_{\Cal H} \otimes S} && \Cal H\otimes \Cal H \ar[ur]^{m}& }
}
\begin{flushleft}
Autrement dit, pour tout $x \in \Cal H$ :$$ \sum_{(x)} S (x_{1}) x_{2}= \varepsilon (x) \un = \sum_{(x)} x_{1}S (x_{2}).$$
\end{flushleft} 
\end{definition}
\begin{proposition} \textbf{( \cite{Sw69} )}
Soit $( \Cal H, m, u,\Delta, \varepsilon, S)$ une alg\`ebre de Hopf, alors on a :
\begin{enumerate}
\item $ S \circ u = u $ et $ \varepsilon \circ S = \varepsilon$.
\item S est un antimorphisme d'alg\`ebres et un antimorphisme de cog\`ebres, i.e. si $\tau$ est la volte on a :
$$m \circ ( S \otimes S ) \circ \tau = S \circ m, \;\;\;\;\;\;\;\; \tau \circ ( S \otimes S ) \circ \Delta = \Delta \circ S.$$
\item Si $\Cal H$ est commutative ou cocommutative, alors $ S^2 = Id_{\Cal H}$.
\end{enumerate}
\end{proposition}
\begin{definition}
Soit $\Cal H$ une alg\`ebre de Hopf et $\Cal A$ une alg\`ebre commutative. On dit que $\varphi : \Cal H \rightarrow \Cal A$ est un caract\`ere si $\varphi (\un)=\un$ et pour tout $x$, $y \in \Cal H$ on a : $\varphi (xy) = \varphi(x) \varphi(y)$, et on dit que $\varphi : \Cal H \rightarrow \Cal A$ est un caract\`ere infinit\'esimal si $\varphi (\un)=0$ et pour tout $x$, $y \in \Cal H$ on a : $$\varphi (xy) = \varphi(x) e (y) + e(x) \varphi(y),$$
o\`u $ e = u \circ \varepsilon $.
\end{definition}
\subsection {Alg\`ebres de Hopf gradu\'ees connexes}
On suppose pour toute la suite que le corps $k$ est de caract\'eristique z\'ero. Une alg\`ebre de Hopf gradu\'ee sur $k$ est un $k$-espace vectoriel gradu\'e :$$\Cal H = \bigoplus_{ n \geq 0} \Cal H _n$$
muni d'un produit $ m : \Cal H \otimes \Cal H \longrightarrow \Cal H$, un coproduit $\Delta : \Cal H \longrightarrow \Cal H \otimes \Cal H $, le tout verifiant les axiomes d'une alg\`ebre de Hopf \cite{Sw69}, et tel que :
$$ m ( \Cal H_p \otimes \Cal H_q ) \subset \Cal H _{p+q} ,$$
$$\Delta ( \Cal H_n ) \subset \bigoplus_{ p+ q = n} \Cal H_p \otimes \Cal H_q ,$$
$$ S ( \Cal H_n ) \subset \Cal H_n .$$
Une alg\`ebre de Hopf gradu\'ee $\Cal H$ sur $k$ est dite connexe si sa partie homog\`ene de degr\'e z\'ero est de dimension un, c'est-\`a-dire r\'eduite \`a $k.\textbf{1}$, o\`u $\un = u ( 1 ) $ d\'esigne l'unit\'e. La donn\'ee d'une telle alg\`ebre de Hopf $ \Cal H$, lorsqu'elle est de plus commutative, \'equivaut \`a la donn\'ee du sch\'ema en groupes pro-nilpotents qui \`a toute alg\`ebre commutative unitaire $ \Cal A $ associe le groupe $ G_\Cal A $ des caract\`eres de $ \Cal H$ \`a valeurs dans $ \Cal A $. Le th\'eor\`eme de Cartier-Milnor-Moore permet de r\'ecup\'erer l'alg\`ebre de Hopf $ \Cal H $ comme le dual gradu\'e de l'alg\`ebre enveloppante $ U(\mathfrak g_k ) $ o\`u $ \mathfrak g_k$ est l'alg\`ebre de Lie du groupe $ G_k $, qui peut se voir comme l'ensemble des caract\`eres infinit\'esimaux de $ \Cal H $ \`a valeurs dans $k$. Les alg\`ebres de Hopf gradu\'ees connexes (commutatives ou non) sont particuli\`erement bien adapt\'ees aux raisonnements par r\'ecurrence sur le degr\'e. Cela vient du fait que pour tout \'el\'ement $x$ homog\`ene de degr\'e $n$ dans $ \Cal H $ on peut \'ecrire en utilisant la notation de Sweedler :
$$\Delta x = x \otimes \textbf{1} + \textbf{1} \otimes x + \sum_{( x ) } x' \otimes x''$$
o\`u les $ x'$ et $ x''$ sont homog\`enes de degr\'e compris entre 1 et $ n-1$. En particulier l'antipode est donn\'e gratuitement par l'une des deux formules de r\'ecurrence ci-dessous :
\begin{equation}
S(x) = - x - \sum_{( x ) } S (x') x''
\end{equation}
\begin{equation}
S(x) = - x - \sum_{( x ) } x' S (x'').
\end{equation}
D. Kreimer a le premier observ\'e que les graphes de Feynman d'une th\'eorie quantique des champs donn\'ee s'organisent en une alg\`ebre de Hopf commutative gradu\'ee connexe \cite{DK98}. Les r\`egles de Feynman r\'egularis\'ees fournissent un caract\`ere de cette alg\`ebre de Hopf \`a valeurs dans une alg\`ebre de fonctions, par exemple l'alg\`ebre des fonctions m\'eromorphes d'une variable complexe dans le cas de la r\'egularisation dimensionnelle.\\
Nous pouvons maintenant expliquer comment renormaliser un caract\`ere $ \varphi$ d'une alg\`ebre de Hopf gradu\'ee connexe : Il faut pour cela que $ \varphi$ soit \`a valeurs dans une alg\`ebre commutative unitaire $\Cal A $ munie d'un sch\'ema de renormalisation, c'est-\`a-dire d'une d\'ecomposition :
\begin{equation}
\Cal A = \Cal A_- \oplus \Cal A_+,
\end{equation}
o\`u $ \Cal A_- $ et $\Cal A_+ $ sont deux sous-alg\`ebres de $\Cal A $, avec $\textbf{1}_\Cal A \in \Cal A_+$. Le sch\'ema minimal \'evoqu\'e plus haut correspond au cas o\`u $ \Cal A$ est l'alg\`ebre (sur $k =\mathbb C$) des fonctions m\'eromorphes d'une variable, $\Cal A_+ $ est la sous-alg\`ebre des fonctions qui sont holomorphes en un $z_0$ fix\'e, et $ \Cal A_- $ est la sous-alg\`ebre des polyn\^omes en $(z - z_0)^{-1}$ sans terme constant. L'espace des applications lin\'eaires de $\Cal H $ dans $\Cal A $ est muni du produit de convolution, donn\'e par : 
\begin{equation}
\varphi \ast \psi = m_\Cal A \circ ( \varphi \otimes \psi ) \circ \Delta.
\end{equation}
Il est facile de v\'erifier que l'espace des caract\`eres de $\Cal H $ \`a valeurs dans $\Cal A $ est un groupe pour le produit de convolution. L'el\'ement neutre $e$ est donn\'e par $e(\textbf{1}) = \textbf{1}_\Cal A $ et $ e(x) = 0$ si $x$ est homog\`ene de degr\'e $ \geq 1 $. L'inverse est donn\'e par la
composition \`a droite avec l'antipode :
\begin{equation}
\varphi^{\ast -1} = \varphi \circ S.
\end{equation}
Chaque caract\`ere $\varphi$ admet une unique d\'ecomposition de Birkhoff :
\begin{equation}
\varphi = \varphi^{\ast -1} _- \ast \varphi_+
\end{equation}
compatible avec le sch\'ema de renormalisation choisi, c'est-\`a-dire telle que $\varphi_+$
prenne ses valeurs dans $\Cal A_+ $ et telle que $ \varphi_- (x) \in \Cal A_- $ pour tout $x$ homog\`ene de
degr\'e $\geq 1$. Les composantes $\varphi_+$ et $\varphi_- $ sont donn\'ees par des formules r\'ecursives assez simples: si on note $\pi$ la projection sur $\Cal A _-$ parall\`element \`a $\Cal A_+ $, et si on suppose que $ \varphi_- (x) $ et $\varphi_+ (x) $ sont connus pour $ x $ de degr\'e $ k \leq n-1 $, on a alors pour tout $ x \in \Cal H_n $ :
\begin{equation}
\varphi_- (x) = - \pi \left( \varphi (x) + \sum_{( x ) } \varphi_- (x') \varphi (x'')\right), \label{CH1}
\end{equation}
\begin{equation}
\varphi_+ (x) = ( I - \pi ) \left( \varphi (x) + \sum_{( x ) } \varphi_- (x') \varphi (x'')\right). \label{CH2}
\end{equation}
On appelle $ \varphi_+ (x) $ le caract\`ere renormalis\'e et $ \varphi_-(x) $ le caract\`ere des contretermes. Le fait remarquable que les deux composantes $ \varphi_+ $ et $ \varphi_- $ soient encore des caract\`eres de $\Cal H $ \`a valeurs dans $\Cal A $ provient de la propri\'et\'e de Rota-Baxter \cite{k.g.m} v\'erifi\'ee par la projection $ \pi$ :
\begin{equation}
\pi (a ) \pi ( b) = \pi ( \pi ( a) b + a \pi (b ) - a b ) 
\end{equation}
\begin{definition} 
Soit $\Cal H $ une alg\`ebre de Hopf gradu\'ee connexe sur le corps $\mathbb C$ des
complexes, et soit $ \varphi $ un caract\`ere de $\Cal H $ \`a valeurs dans l'alg\`ebre $\Cal A $ des fonctions
m\'eromorphes, munie du sch\'ema de renormalisation minimal en $z_0$. Alors le caract\`ere
\`a valeurs scalaires donn\'e par $ x \longmapsto \varphi_+ (x) (z_0) $ d\'efinit la valeur renormalis\'ee du
caract\`ere $\varphi $ en $z_0$.
\end{definition} 
L'application lin\'eaire $ b ( \varphi ):\Cal H \longrightarrow \Cal A $ donn\'ee par $ b( \varphi )( \textbf{1}) = 0$ et pour tout $x \in \Cal H$ par :
$$ b ( \varphi ) ( x ) = \varphi(x) + \sum_{( x ) } \varphi_- (x') \varphi (x''),$$
est nomm\'ee pr\'eparation de Bogoliubov et s'\'ecrit $ b ( \varphi ) = \varphi_- \ast ( \varphi - e )$. Les formules de r\'ecurrence \eqref {CH1} et \eqref {CH2} s'\'ecrivent de mani\`ere plus compacte :
\begin{eqnarray*}
\varphi_- &=& e + P ( \varphi_- \ast \lambda )\\
&=& e + P (\lambda ) + P ( P (\lambda ) \ast \lambda ) +........+ \underbrace{P ( P (...P (}_{ n fois}\lambda ) \ast \lambda ).... \ast \lambda )+.... \\ 
\end{eqnarray*}
et: 
\begin{eqnarray*}
\varphi_+ &=& e + \wt P ( \varphi_+ \ast \xi )\\
&=& e + \wt P (\xi ) + \wt P ( \wt P (\xi ) \ast \xi ) +........+ \underbrace{\wt P ( \wt P (...\wt P (}_{ n fois}\xi ) \ast \xi ).... \ast \xi )+.... \\ 
\end{eqnarray*}
avec $ \lambda:= e - \varphi $, $ \xi:= e - \varphi^{-1}$, et o\`u $P$ et $\wt P$ sont les projections sur $ L(\Cal H, \Cal A)$ d\'efinies par $ P( \lambda ) = \pi \circ \lambda $ et $\wt P (\xi ) = ( I - \pi ) \circ \xi $, respectivement. A. Connes et D. Kreimer ont montr\'e dans \cite{A.D2000} que lorsque $ \Cal H$ est l'alg\`ebre de Hopf des graphes de Feynman associ\'es \`a une th\'eorie des champs renormalisable, cette d\'efinition de la renormalisation compatible avec l'algorithme BPHZ des physiciens (tel qu'il est expos\'e par exemple dans \cite{J.C84}).
\section {Deux alg\`ebres de Hopf gradu\'ees connexes en interaction}
On se placera dans le cadre suivant : $\Cal H$ et $ \Cal K$ sont deux alg\`ebres de Hopf gradu\'ees connexes commutatives et $\Phi:\Cal H \longrightarrow \Cal K \otimes\Cal H $ une coaction \`a gauche qui est en m\^eme temps un morphisme d'alg\`ebres gradu\'ees, et telle que le diagramme suivant commute :
\diagramme{
\xymatrix{
\Cal H \ar[d]_{\Delta} \ar[rr]^{\Phi} 
& &\Cal K \otimes\Cal H \ar[dd]^{I \otimes \Delta }\\
\Cal H \otimes\Cal H \ar[d]_{\Phi\otimes\Phi}\\
\Cal K \otimes\Cal H\otimes \Cal K \otimes \Cal H \ar[rr]^{m^{13}} 
&& \Cal K\otimes\Cal H\otimes \Cal H }
}
i.e :
\begin{equation}
(\mop{Id}_{\Cal K}\otimes\Delta_{\Cal H})\circ\Phi=m^{1,3}\circ(\Phi\otimes\Phi)\circ\Delta_{\Cal H},\label{th1}
\end{equation}
o\`u\;\; $m^{1,3}:\Cal K \otimes\Cal H\otimes \Cal K \otimes \Cal H
\longrightarrow \Cal K\otimes\Cal H\otimes\Cal H$ \;\; est d\'efini par :
$$
m^{1,3}(a\otimes b\otimes c\otimes d)=ac\otimes b\otimes d,
$$
et $ \Phi $ s'exprime en notation de Sweedler pour tout $ x \in \Cal H$ par :
\begin{equation}
\Phi (x) = \sum_{(x)} x_0 \otimes x_1 = \un_\Cal K \otimes x + \sum_{(x)} x^{(')} \otimes x^{('')},\label{ph1}
\end{equation} 
avec : $ 1 \leq \left| x^{('')} \right| \leq  \left|\; x \;\right| -1 $  et  $  \left| x^{('')} \right| + deg\; x^{(')} = \left| \;x \;\right| $   o\`u  $ \left| ... \right| $ d\'esigne le degr\'e  dans $ \Cal H $ et $ deg $  d\'esigne le degr\'e dans $ \Cal K $. Le cadre ci-dessus est inspir\'e par les deux exemples suivants : 
\subsection{Les arbres enracin\'es} 
D. Calaque, K. Ebrahimi-Fard et D. Manchon ont etudi\'e l'alg\`ebre de Hopf $\Cal H$ de Connes-Kreimer gradu\'ee suivant le nombre de sommets, dans \cite{ckm}, comme comodule sur une alg\`ebre de Hopf $\Cal K$ d'arbres enracin\'es gradu\'ee suivant le nombre d'ar\^etes. Cette structure est d\'efinie de la fa\c con suivante : Pour tout arbre non vide $t$ on a : $$\Phi(t)=\Delta_{\Cal K} (t) = \sum_{s\subseteq t} s \otimes t/s,$$ o\`u la notation $s\subseteq t$ exprime le fait que $s$ est une sous for\^et de l'arbre $t$, c-\`a-d $s$ est soit la for\^et triviale $\racine$, ou une collection $(t_1, \cdots ,t_n)$ de sous-arbres disjoints de $t$, chacun d'eux contenant au moins une ar\^ete. En particulier, deux sous-arbres d'une sous for\^et ne peuvent avoir aucun sommet en commun, et $t/s$ est l'arbre obtenu par contraction des composantes connexes de $s$ en un sommet, et pour $\un$ on a : $\Phi(\un)=\racine \otimes \un$. On peut \'ecrire $\Phi(t)$ encore de la mani\`ere suivante :
\begin{eqnarray*}
\Phi(t) &=& \Delta_{\Cal K} (t) = \sum_{s\subseteq t} s \otimes t/s\\
&=& \racine \otimes t + \left( t \otimes \racine  +  \sum_{s \hbox{ \sevenrm sous-for\^et propre de }t} s \otimes t/s \right ),
\end{eqnarray*}
ce qui montre que la formule \eqref{ph1} est v\'erifi\'ee.\\
Le th\'eor\`eme suivant donne la relation entre cette structure de comodule et le coproduit de Connes-Kreimer $\Delta_{\Cal H}$ d\'efini par :
\begin{equation}
\Delta_{\Cal H}(t) = t \otimes \un + \un \otimes t + \sum_{c \in {\mop{\tiny{Adm}}(t)}} P^c(t)\otimes R^c(t),
 \end{equation}
o\`u $\mop{Adm}(t)$ designe l'ensemble des coupes admissibles d'une for\^et $ t$ (rappelons qu'une  coupe admissible de $t$ est une coupe non vide tel que tout trajet d'un sommet de $t$ vers un autre ne rencontre au plus qu'une seule coupe \'el\'ementaire). Une coupe admissible envoie $t$ vers un couple $( P^c(t) , R^c(t) )$ telle que $R^c(t)$ est la composante connexe de la racine de $t$ apr\`es la coupe, et $ P^c(t)$ est la for\^et form\'ee par les autres composantes connexes. (Voir \cite{ckm} et \cite{lf}). 
\begin{theorem}\cite{ckm} 
L'identit\'e suivante  est v\'erifi\'ee :
\begin{equation}\label{codistrib}
(\mop{Id}_{\Cal K} \otimes\Delta_{\Cal H })\circ\Phi=m^{1,3}\circ(\Phi\otimes\Phi)\circ\Delta_{\Cal H},
\end{equation}
o\`u $ m^{1,3}:\Cal K \otimes \Cal H \otimes \Cal K \otimes \Cal H 
\longrightarrow \Cal K \otimes \Cal H \otimes \Cal H $ est d\'efinie par :
\begin{equation}
m^{1,3}(a\otimes b\otimes c\otimes d)=ac\otimes b\otimes d.
\end{equation}
\end{theorem}
\subsection{Les graphes de Feynman orient\'es sans cycle} 
Un graphe de Feynman orient\'e est un graphe orient\'e (non plan) avec un nombre fini de sommets et d'ar\^etes, qui peuvent \^etre internes ou externes. Une ar\^ete interne est une ar\^ete connect\'ee aux deux extr\'emit\'es \`a un sommet, une ar\^ete externe est une ar\^ete avec une extr\'emit\'e ouverte, l'autre extr\'emit\'e \'etant reli\'ee \`a un sommet.\\

Un cycle dans un graphe de Feynman orient\'e est une collection finie d'ar\^etes orient\'ees internes $(e_1,..., e_n)$ tels que le but de $e_k$ coincide avec la source de $e_{k+1}$ pour tout $k = 1,. . . , n$ modulo $n$.\\

Pour toute partie non vide $P$ de l'ensemble $\Cal V (\Gamma)$ des sommets de $\Gamma$, le sous graphe $\Gamma (P)$ est d\'efini comme suit : les ar\^etes internes de $\Gamma (P)$ sont les ar\^etes internes de $\Gamma$ avec source et but dans $ P$, et les ar\^etes externes sont les ar\^etes externes de $\Gamma$ avec la source ou le but dans $P$, ainsi que les ar\^etes internes de $\Gamma$ avec une extr\'emit\'e dans $P$ et l'autre extr\'emit\'e hors de $P$.\\

Un sous-graphe couvrant de $\Gamma$ est un graphe de Feynman orient\'e $\gamma$ (pas forcement connexe), donn\'e par une collection ${\Gamma(P_1),. . . , \Gamma(P_n)}$ de sous-graphes connexes telle que $P_j \cap P_k = \emptyset $ pour $j \neq k$, et telle que tout sommet de $\Gamma$ appartient \`a un certain $P_j$ pour $j \in \{1,. . . , n\}$.\\

Pour tout sous-graphe couvrant $\gamma$, le graphe contract\'e $\Gamma / \gamma$ est d\'efini par contraction de toutes les composantes connexes de $\gamma$ sur un point. On dira qu'un sous-graphe couvrant $\gamma$ de $\Gamma$ est compatible avec l'ordre partiel si le graphe contract\'e $\Gamma / \gamma$ est sans cycle.\\

Les graphes de Feynman orient\'es sans cycles engendrent \`a la fois une alg\`ebre de Hopf $\Cal H$ (gradu\'ee suivant le nombre des sommets) et une big\`ebre $\wt {\Cal K}$ (gradu\'ee suivant le nombre des ar\^etes internes) \cite{Dm11}. La coaction \`a gauche de $\wt {\Cal K}$ sur $\Cal H$ est d\'efinie par :
$$ \wt {\Phi}(\un_{\Cal H}) = \un_{\wt{\Cal K}} \otimes \un_{\Cal H},$$
et pour tout graphe non vide $\Gamma$ par :
\begin{eqnarray*}
\wt {\Phi}(\Gamma) &=& \Delta_{\wt{\Cal K}} (\Gamma) = \sum_{{\gamma \hbox{ \sevenrm sous graphe couvrant de } \Gamma \atop \hbox{ \sevenrm compatible avec l'ordre partiel}}} \gamma \otimes \Gamma / \gamma.
\end{eqnarray*}
Le coproduit sur $\Cal H $ est d\'efini pour tout graphe orient\'e sans cycle $\Gamma$ par :
\begin{equation}
\Delta_{\Cal H}(\Gamma) =  \sum_{V_1\cup V_2 = \Cal V (\Gamma),V_1\prec V_2} \Gamma (V_1) \otimes \Gamma (V_2)  ,
\end{equation}
o\`u l'in\'egalit\'e $V_1\prec V_2$ signifie que pour tout $v_1 \in V_1$ et $v_2 \in V_2$ avec $v_1$ et $v_2$ comparable, on a $v_1\prec v_2$ dans l'ensemble partiellement ordonn\'e des sommets de $\Gamma$ not\'e $\Cal V (\Gamma)$.\\
D'apr\`es \cite[Theorem 2]{Dm11} le coproduit $\Delta_{\Cal H}$ et la coaction $\wt{\Phi}$ v\'erifient bien : 
\begin{equation}
(\mop{Id}_{\wt{\Cal K}}\otimes\Delta_{\Cal H})\circ \wt{\Phi} = m^{1,3}\circ(\wt{\Phi} \otimes \wt{\Phi})\circ\Delta_{\Cal H}.
\end{equation}
En quotientant par l'id\'eal engendr\'e par les \'el\'ements $\Gamma - \un_{\wt{\Cal K}}$ o\`u $\Gamma$ est un graphe sans ar\^etes internes, la big\`ebre ${\wt{\Cal K}}$ donne naissance \`a une alg\`ebre de Hopf gradu\'ee connexe $\Cal K$, et la coaction $\Phi$ d\'eduite de $\wt{\Phi}$ par passage au quotient v\'erifie la formule \eqref{th1}. La coaction $\Phi$ s'\'ecrit alors :
\begin{eqnarray*}
\Phi(\Gamma) &=& \un_{\wt{\Cal K}} \otimes \Gamma + \left( \Gamma \otimes \racine  +  \sum_{{\gamma \hbox{ \sevenrm sous graphe couvrant propre de } \Gamma \atop \hbox{ \sevenrm compatible avec l'ordre partiel}}} \gamma \otimes \Gamma / \gamma \right ),
\end{eqnarray*}
et verifie bien la formule \eqref{ph1}.
\subsection {Groupes de caract\`eres}
On rappelle ici que $ \Cal A$ est l'alg\`ebre (sur $ k =\mathbb C $) des fonctions m\'eromorphes d'une variable, $\Cal A_+ $ est la sous-alg\`ebre des fonctions qui sont holomorphes en un $z_0$ fix\'e, et $ \Cal A_- $ est la sous-alg\`ebre des polyn\^omes en $(z - z_0)^{-1}$ sans terme constant. On d\'esigne par $ G_\Cal A$ (resp. $G^\Cal K_\Cal A$) le groupe des caract\`eres de $\Cal H$ (resp. de $\Cal K$) \`a valeurs dans $\Cal A \;, G_{\Cal A_+}$ (resp. $ G^\Cal K_{\Cal A_+}$ ) le groupe des caract\`eres de $\Cal H$ (resp. de $\Cal K$) \`a valeurs dans $\Cal A_+$ et $ G_c$ (resp. $G^\Cal K_c$) le groupe des caract\`eres de $\Cal H$ (resp. de $\Cal K$) \`a valeurs constants.
\begin{remark}
Dans toute la suite on utilise des notations similaires ($ \mathfrak g_{\Cal A}$, $\mathfrak g^\Cal K_{\Cal A}$, $\mathfrak g_{\Cal A_+}$, $\mathfrak g^\Cal K_{\Cal A_+}$, $\mathfrak g_c$, $\mathfrak g^\Cal K_c $) pour les alg\`ebres de Lie des caract\`eres infinit\'esimaux associ\'ees aux groupes des caract\`eres.
\end{remark}
Tout $ \alpha  \in  G^\Cal K_{\Cal A_+} $ s'\'ecrit sous la forme ${ \exp}^{\star} X$ o\`u $ X \in \mathfrak g^\Cal K_{\Cal A}$. On d\'efinit la bijection $Z$ par :
\begin{eqnarray*}
Z:G^\Cal K_\Cal A &\longrightarrow& G^\Cal K_\Cal A \\
{ \exp}^{\star} X &\longmapsto&{ \exp}^{\star} z X
\end{eqnarray*} 
o\`u ${ \exp}^{\star} z X$ est d\'efini par : 
 $${ \exp}^{\star} z X ( x) = \sum_{n \geq 0 } \frac{z^n}{n !}X^{\star n} ( x ),$$
pour tout $ x \in \Cal K $. L'inverse de $ Z$ est donn\'e par la formule suivante : $$ Z^{-1} ( { \exp}^{\star} X )(x) = { \exp}^{\star} z^{-1} X (x) = \sum_{n \geq 0 } \frac{z^{- n}}{n !}X^{\star n}.$$
\begin{remark} La somme pr\'ec\'edente est finie car elle s'arr\^ete \`a $ n = \left| x \right| $ o\`u $ \left| x \right| $ d\'esigne le degr\'e de $x$.
\end{remark}
Pour tout $ g, g' \in G^\Cal K_\Cal A$ on pose : $$ g \starz g':= Z^{-1} ( Z( g ) \star Z( g' )).$$
\begin{definition}
L'action de $ G^\Cal K_{\Cal A}$ sur $ G_\Cal A$ est d\'efinie pour tout $ g \in G^\Cal K_\Cal A , \varphi \in G_\Cal A , x \in \Cal H$ et $z \in \mathbb C$ par :
$$ ( g \starz \varphi) ( x ) ( z ) := ( Z( g ) \star \varphi) ( x ) ( z ).$$
\end{definition}
Cette formule d\'efinit bien une action. En effet pour tout $ g , g' \in G^\Cal K_\Cal A , \varphi \in G_\Cal A$ on a :
\begin{eqnarray*}
g \starz ( g' \starz \varphi)&=& Z ( g ) \star ( g' \starz \varphi)\\
&=& Z ( g ) \star ( Z ( g' ) \star \varphi)\\
&=& ( Z ( g ) \star Z ( g' )) \star \varphi\\
&=& Z ( g \starz g' ) \star \varphi\\
&=& ( g \starz g' ) \starz \varphi.
\end{eqnarray*}

\begin{proposition} \label{transposeL}
Soit $\alpha :\Cal K \longrightarrow \Cal A_+$ une transformation lin\'eaire. L'application $B_\alpha : \Cal H \longrightarrow \Cal H$ d\'efinie par :
$$ B_\alpha = (\alpha \otimes \mop{Id}_{\Cal H}) \circ \Phi $$ i.e :
\begin{equation}
B_\alpha(x)=\sum_{(x)}<\alpha,\,x_1>x_0
\end{equation}
satisfait l'identit\'e :
\begin{equation}
\Delta_{\Cal H}\circ{}B_\alpha={}B_{{}^t\!m \alpha}\circ\Delta_{\Cal H},
\end{equation}
o\`u  ${}^t\!m : \Cal K^*\to (\Cal K\otimes\Cal K)^*$ est d\'efini par :
${}^t\!m(\alpha)(x\otimes y) := \alpha(xy)$ et 
\begin{equation*}
B_{{}^t\!m\alpha}:=({}^t\!m\alpha\otimes
\mop{Id}_{\Cal H}\otimes
\mop{Id}_{\Cal H})\circ\tau_{2,3}\circ(\Phi\otimes\Phi).
\end{equation*}
En particulier si $\alpha\in\Cal K^\circ$ alors
${}^t\!m \alpha=\sum_{(\alpha)}\alpha_1\otimes \alpha_2\in\Cal K^\circ\otimes\Cal K^\circ$ et :
\begin{equation}
\Delta_{\Cal H} \circ{} B_\alpha=\sum_{(\alpha)}({} B_{\alpha_1}\otimes{} B_{\alpha_2})\circ \Delta_{\Cal H}.
\end{equation}
\end{proposition} 

\begin{proof}[Preuve]
L'op\'erateur $ B_\alpha$ est la transpos\'ee de l'op\'erateur de multiplication \`a gauche 
$$ L_\alpha:\Cal H^\circ\to{\Cal H}^\circ$$
donn\'ee par la structure de $\Cal H^\circ$-module \`a gauche (i.e. $L_\alpha(b)=\alpha\star b$). De m\^eme, si $\alpha\in\Cal K^\circ$ alors ${}^t\!m\alpha\in \Cal K^\circ\otimes \Cal K^\circ$, et $B_{{}^t\!m\alpha}$ est la transpos\'ee de l'op\'erateur de multiplication \`a gauche$$ L_{{}^t\!m\alpha} :
\Cal H^\circ \otimes
\Cal H^\circ \to
\Cal H^\circ \otimes
\Cal H^\circ.$$
La structure de $\Cal H^\circ \otimes \Cal H^\circ$-module \`a gauche donn\'ee par la transpos\'ee de $\wt\Phi=\tau_{2,3}\circ(\Phi\otimes\Phi)$, o\`u la notation $\tau_{2,3}$ d\'esigne la permutation des deux termes interm\'ediaires, i.e : $$\tau_{2,3}(a\otimes b\otimes c\otimes d)=a\otimes c\otimes b\otimes d.$$
La preuve de la proposition est un calcul direct bas\'e sur la d\'efinition de la coaction $\Phi$ :
\allowdisplaybreaks{
\begin{eqnarray*}
\Delta_{ \Cal H} \circ B_\alpha&=&\Delta_{\Cal H}\circ (\alpha\otimes\mop{Id})\circ\Phi\\
&=&(\alpha\otimes\mop{Id}\otimes\mop{Id})\circ (\mop{Id}\otimes\Delta_{ \Cal H})\circ\Phi\\
&=&(\alpha\otimes\mop{Id}\otimes\mop{Id})\circ m^{1,3}\circ(\Phi\otimes\Phi)\circ\Delta_{\Cal H}\\
&=&({}^t\!m\alpha\otimes \mop{Id}\otimes \mop{Id})\circ \tau_{2,3}\circ(\Phi\otimes\Phi)\circ\Delta_{ \Cal H}\\
&=& B_{{}^t\!m \alpha}\circ\Delta_{ \Cal H}.
\end{eqnarray*}}
\end{proof}

\begin{proposition} 
Si $ \alpha :\Cal K \longrightarrow \Cal A $ est un caract\`ere infinit\'esimal de $ \Cal K$ alors l'op\'erateur $ B_\alpha $ est une bid\'erivation de l'alg\`ebre de Hopf $ \Cal H$.
\end{proposition}
\begin{proof}[Preuve]
Le fait que $ B_\alpha $ est une cod\'erivation d\'ecoule imm\'ediatement de la proposition \ref{transposeL} et du fait que $\alpha $ est infinit\'esimal. Montrons maintenant que $B_\alpha$ est une d\'erivation. Soient $x , y \in \Cal H$ :
\begin{eqnarray*}
B_\alpha ( x y ) &=& \sum_{(x y)}< \alpha ,\,( x y)_1> ( x y)_0\\
&=&\sum_{(x )} \sum_{( y)}<\alpha,\, x_1 y_1> x_0 y_0 \\
&=& \sum_{(x )} \sum_{( y)}( <\alpha ,\, x_1 >e(y_1)+ <\alpha,\, y_1 >e(x_1) )x_0 y_0 \\
&=& \sum_{(x )} <\alpha ,\, x_1 >x_0 y_0+ \sum_{( y)} <\alpha ,\, y_1 > x_0 y_0 \\
&=& B_\alpha ( x ) y + x B_\alpha ( y ). 
\end{eqnarray*}
Ce qui prouve que $B_\alpha $ est une d\'erivation.
\end{proof}
\begin{corollary} 
Si $ \alpha :\Cal K \longrightarrow \Cal A $ est un caract\`ere infinit\'esimal de $ \Cal K$, alors $\varphi \longmapsto \varphi \circ B_\alpha = \alpha \star \varphi $ est une d\'erivation de l'alg\`ebre de Hopf $ \Cal L ( \Cal H, \Cal A)$ pour le produit de convolution.
\end{corollary}
\begin{proof}[Preuve]
Soient $\varphi, \psi \in G_\Cal A$.$$ \alpha \star ( \varphi \ast \psi) = \sum_{(\alpha)} ( \alpha_{1}\star\varphi ) \ast ( \alpha_{2}\star\psi ).$$
Comme $\alpha$ est un caract\`ere infinit\'esimal, alors il est primitif pour ${}^t\!m$. On a donc : $$ {}^t\!m ( \alpha ) = \alpha \otimes 1_{\Cal K^\circ} + 1_{\Cal K^\circ} \otimes \alpha. $$
D'o\`u : $ \alpha \star ( \varphi \ast \psi) = ( \alpha \star \varphi ) \ast \psi + \varphi \ast ( \alpha \star \psi).$
\end{proof}
\section{Groupe de renormalisation}
Pour tout $\alpha \in \mathfrak g^{\Cal A_ +}_ \Cal K$, on obtient un groupe \`a un param\`etre $\theta_{t,\alpha}$ d'automorphismes de $ G_\Cal A$ d\'efini pour tout $\varphi \in G_\Cal A$ par :
\begin{equation}
\theta_{t,\alpha } ( \varphi )( x ) ( z ) = ({ exp}^{\star} t z \alpha \star \varphi ) (x ) (z ). \label{CHV1}
\end{equation}
La formule \eqref{CHV1} d\'efinit \'egalement un sous-groupe \`a un param\`etre d'automorphismes de l'alg\`ebre $(\Cal L ( \Cal H, \Cal A) , \ast) $.
On note :
\begin{equation}
\varphi_{t,\alpha} := \theta_{t,\alpha} ( \varphi ).
\end{equation}
En termes de d\'ecomposition de Birkhoff $ \varphi_{t,\alpha}$ s'\'ecrit :
$$ \varphi_{t,\alpha} = (\varphi_{t,\alpha})^{\ast-1}_- \ast (\varphi_{t,\alpha})_+ .$$
On note $ G^\alpha_\Cal A $ l'ensemble des caract\`eres $ \varphi$ de $ \Cal H$ \`a valeurs dans $ \Cal A$ qui v\'erifient :
$$ \frac{d}{dt} (\varphi_{t,\alpha})_- = 0.$$
\begin{proposition} 
Pour tout $ \alpha \in \mathfrak g^{\Cal A_ +}_ \Cal K$, l'\'equation :
\begin{equation}
\alpha \star \varphi = \varphi\ast \gamma \label{CHV2}
\end{equation}
d\'efinit une application :
\begin{eqnarray*}
\wt{\Cal R}_\alpha: G_\Cal A &\longrightarrow& \mathfrak g_\Cal A \\
\varphi &\longmapsto& \gamma
\end{eqnarray*} 
\end{proposition}
\begin{proof}[Preuve]
Pour $ x= \un_\Cal H $ : $\gamma ( \un_\Cal H ) =0$ et pour $\left|{x}\right| = 1$ l'\'equation \eqref{CHV2} s'\'ecrit en utilisant la notation de Sweedler :
\begin{eqnarray*}
\alpha( \un_\Cal K ) \varphi ( x ) = \varphi ( x ) \gamma ( \un_\Cal H ) + \varphi ( \un_\Cal H ) \gamma ( x ).
\end{eqnarray*}
Le fait que $ \alpha( \un_\Cal K ) = \gamma ( \un_\Cal H ) =0$ implique que $\gamma ( x ) = 0$ et pour $ x \in \mop{Ker} \varepsilon$, l'\'equation \eqref{CHV2} s'\'ecrit en utilisant la notation de Sweedler :
\begin{eqnarray*}
\alpha( \un_\Cal K) \varphi (x ) + \sum_{ x } \alpha( x^{(')}) \varphi ( x^{('')}) = \gamma ( x ) + \sum_{ (x)} \varphi ( x') \gamma ( x''). 
\end{eqnarray*} 
\begin{eqnarray*}
\text{ Donc : } \gamma ( x ) = \sum_{ x } \alpha(x^{(')}) \varphi ( x^{('')}) - \sum_{ (x) } \varphi ( x') \gamma ( x''). 
\end{eqnarray*}
Ce qui nous permet de d\'efinir $ \gamma ( x )$ par r\'ecurrence sur le d\'egr\'e de $x''$. 
Soient $ \varphi \in G_\Cal A$ \;et $ x , y \in \Cal H$.
\begin{eqnarray*}
\gamma ( xy )&=& \varphi^{\ast-1} \ast (\alpha \star \varphi)(xy)\\
&=& \varphi^{\ast-1} \ast ( \varphi \circ B_\alpha) (x y)\\
&=&\sum_{(xy )}\varphi^{\ast-1}( (x y)_1) \varphi \circ B_\alpha ( (x y)_2) \\
&=& \sum_{(x )(y)} \varphi^{\ast-1}( x_1 y_1) \varphi \circ B_\alpha( x_2 y_2) \\
&=& \sum_{(x )(y)} \varphi^{\ast-1}( x_1) \varphi^{\ast-1}(y_1) \varphi ( B_\alpha)( x_2) y_2 + x_2 B_\alpha( y_2)) \\
&=&\sum_{(x )(y)} \varphi^{\ast-1}( x_1) \varphi^{\ast-1}(y_1) ( \varphi \circ B_\alpha( x_2)\varphi ( y_2 ) + \varphi( x_2 ) \varphi \circ B_\alpha( y_2))\\
&=&\varphi^{\ast-1} \ast ( \varphi \circ B_\alpha) ( x ) e ( y ) + e ( x ) \varphi^{\ast-1} \ast ( \varphi \circ B_\alpha) ( y ) \\
&=&\gamma ( x ) e ( y ) +e ( x ) \gamma ( y ). 
\end{eqnarray*}
D'o\`u $ \gamma$ est un caract\`ere infinit\'esimal. 
\end{proof}
\begin{proposition} 
Soit : \;\;\;\begin{eqnarray*}
{\Cal R}_\alpha: \mathfrak g_\Cal A &\longrightarrow& \mathfrak g_\Cal A \\ a &\longmapsto& \gamma
\end{eqnarray*} 
\text{ Alors on a : } \;\;\; \;\;\;\;\;\;\;\; \;\; $ \wt{\Cal R}_\alpha ( \varphi ) = \varphi^{\ast-1} \ast ( \alpha \star \varphi) \;\;\; \; \text{ et } \;\;\;\;{\Cal R}_\alpha ( a ) = e^{\ast- a} \ast ( \alpha \star e^{\ast a} ) $.
\end{proposition} 
\begin{proof}[Preuve]
$ \wt{\Cal R}_\alpha ( \varphi ) = \gamma$ alors en utilisant l'\'equation \eqref{CHV2} on ontient : $\alpha \star \varphi = \varphi \ast \gamma$, d'o\`u : 
$$ \wt{\Cal R}_\alpha ( \varphi ) = \varphi^{\ast-1} \ast ( \alpha \star \varphi)$$
Comme : ${\Cal R}_\alpha = \wt{\Cal R}_\alpha \circ exp$ on obtient imm\'ediatement d'apr\`es le r\'esultat ci dessus : $$ {\Cal R}_\alpha ( a ) = e^{\ast- a} \ast (\alpha \star e^{\ast a} ). $$
\end{proof}

\begin{remark}
Si $ \alpha = 0$ alors on a :  $ \theta_{t,\alpha } ( \varphi ) = \varphi $ et  $\wt{\Cal R}_\alpha \equiv 0.$  
\end{remark}
\subsection{L'op\'erateur $ E_\alpha $}
Soit $ \Cal H$ une alg\`ebre de Hopf gradu\'ee connexe, pour tout $ \alpha \in \mathfrak g^{\Cal A_ +}_ \Cal K$ on d\'efinit l'op\'erateur $ E_\alpha $ par :
$$ E_\alpha:= S\ast B_\alpha, $$
o\`u $S$ est l'antipode de $ \Cal H$.
\begin{proposition} 
Si $ \Cal H$ une alg\`ebre de Hopf gradu\'ee connexe commutative la correspondance $ \wt{\Cal R}_\alpha $ se r\'eduit \`a la composition \`a droite avec $ E_\alpha $, \;\;\;\;\; i.e : \;\;\;\;\; $ \wt{\Cal R}_\alpha (\varphi ) = \varphi \circ E_\alpha $.
\end{proposition}
\begin{proof}[Preuve]
\begin{eqnarray*}
\varphi \circ E_\alpha &=& \varphi \circ ( S \ast B_\alpha)\\
&=& (\varphi \circ S )\ast (\varphi \circ B_\alpha)\\
&=& \varphi^ {\ast-1}\ast (\alpha \star \varphi )\\
&=& \wt{\Cal R}_\alpha (\varphi ).
\end{eqnarray*}
\end{proof}
\begin{proposition} 
$$ {\Cal R}_\alpha ( a ) =\int_{0}^{1} e^{\ast- sa} \ast ( a \circ B_\alpha ) \ast e^{\ast sa} ds = \frac{ 1 - e^{- ad\; a} }{ ad\; a }. ( a \circ B_\alpha ).$$
\end{proposition}
\begin{proof}[Preuve]
Pour tout $ u \in \mathbb C $ nous avons :
$$ e^{\ast ua} \circ B_\alpha = e^{\ast ua} \ast {\Cal R}_\alpha ( u a ).$$
On pose $ u = t + s $ et on utilise la propri\'et\'e de groupe $ e^{\ast (t + s )a} = e^{\ast ta} \ast e^{\ast sa}$. En utilisant la propri\'et\'e de d\'erivation :
$$ ( e^{\ast ta} \ast e^{\ast sa} ) \circ B_\alpha = ( e^{\ast ta} \circ B_\alpha ) \ast e^{\ast sa} + e^{\ast ta} \ast ( e^{\ast sa} \circ B_\alpha ).$$
On trouve :
$$ e^{\ast ( t + s )a} \circ B_\alpha = e^{\ast ( t + s )a} \ast ( {\Cal R}_\alpha ( s a ) + e^{\ast- sa} \ast {\Cal R}_\alpha ( t a ) \ast e^{\ast sa} ). $$
En posant\;\; $\gamma ( t ) = {\Cal R}_\alpha ( t a )$ : L'\'equation pr\'ec\'edente devient :
$$ \gamma ( t +s ) = \gamma ( s ) + e^{\ast- sa} \ast \gamma ( t ) \ast e^{\ast sa}.$$
Nous avons $ \gamma ( 0 ) = 0$. En d\'erivant l'\'equation pr\'ec\'edente par rapport \`a $ s $ et prenant $ s = 0$, on obtient :
$$\dot{\gamma} (t ) = \dot{\gamma} (0 ) + [ \gamma ( t ) , a ].$$
On d\'erive l'\'equation pr\'ec\'edente par rapport \`a $ t$ on a :
$$\ddot{\gamma} (t ) = [ \dot{\gamma} ( t ) , a ]. $$
La solution de cette \'equation diff\'erentielle du premier ordre est donn\'ee par :
$$\dot{\gamma} (t ) = e^{\ast - ta} \ast \dot{\gamma} ( 0 ) \ast e^{\ast ta}.$$
En d\'erivant l'egalit\'e $ e^{\ast ta} \circ B_\alpha = e^{\ast ta} \ast \gamma (t )$ \`a $ t = 0$ on obtient imm\'ediatement :
$$ \dot{\gamma} ( 0 ) = a \circ B_\alpha.$$
On int\`egre puis on prend $t =1$, ce qui prouve la proposition. 
\end{proof}
\subsection{Fonction $ \beta_\alpha$} 
On d\'esigne par $ G^\alpha_{\Cal A_-} $ l'ensemble des \'el\'ements $\varphi \in G^\alpha_\Cal A $ tels que $ \varphi = \varphi^{\ast-1}_-$. Comme la composition \`a droite avec $ B_\alpha$ est une d\'erivation pour le produit de convolution, l'application $ \wt{\Cal R}_\alpha $ v\'erifie la propri\'et\'e de cocycle :
\begin{equation}
\wt{\Cal R}_\alpha (\varphi \ast \psi) = \wt{\Cal R}_\alpha ( \psi) + \psi^{\ast_-1} \ast \wt{\Cal R}_\alpha (\varphi )\ast \psi. \label{CHV3}
\end{equation}
\begin{definition}
Pour toute $\varphi \in \Cal L (\Cal H, \Cal A)$, nous associons une forme lin\'eaire $ \mop {Res} \varphi$ sur $\Cal H$ par extraction du coefficient de $z^{-1}$: plus pr\'ecis\'ement, si nous avons pour tout $x \in \Cal H$ et pour tout $z$ dans un voisinage de $0$ :
$$\varphi(x)(z) = \sum_{n=-N}^{+\infty} \varphi_n (x) z^n,$$
avec $\varphi_n (x) \in \mathbb C$, alors : 
$$ \mop {Res} \varphi (x) := \varphi_{-1} (x).$$
\end{definition}
\begin{theorem}
\begin{enumerate}
\item Pour tout $ \varphi \in G_\Cal A$ il y a une famille \`a un param\`etre $ h_{t, \alpha}$ dans $ G_\Cal A$ telle que : $\varphi_{t,\alpha} = \varphi \ast h_{t, \alpha}$ et on a :
\begin{equation}
\dot h_{t, \alpha} = \frac{d}{dt} h_{t,\alpha} = h_{t,\alpha} \ast z \wt{\Cal R}_\alpha ( h_{t,\alpha}) +z \wt{\Cal R}_\alpha ( \varphi)\ast h_{t,\alpha}. \label{CHV4}
\end{equation}
\item$ z \wt{\Cal R}_\alpha$ se restreint en une application de $ G^\alpha_{\Cal A} $ dans $ \mathfrak g^{\Cal A} \cap \Cal L (\Cal H, \Cal A_+)$. Par ailleurs $ z \wt{\Cal R}_\alpha$ envoie $ G^\alpha_{\Cal A_-} $ sur l'ensemble des \'el\'ements de $ \mathfrak g^{\Cal A} $ \`a valeurs constantes.
\item Pour tout $ \varphi \in G^\alpha_{\Cal A}$, le terme constant de $ h_{t,\alpha}$ defini par :
\begin{eqnarray*}
F_{t, \alpha }(\varphi)(x) = \lim_{z\longrightarrow0} h_{t,\alpha} (x) (z)
\end{eqnarray*}
est un sous-groupe \`a un param\`etre de $ G_{\Cal A} \cap \Cal L (\Cal H , \mathbb C).$
\end{enumerate}
\end{theorem}
\begin{proof}[Preuve]
\begin{enumerate}
\item Pour tout $ \varphi \in G_\Cal A$ on \'ecrit : $\varphi_{t,\alpha} = \varphi \ast h_{t, \alpha}$ pour tout $ h_{t, \alpha}$ dans $ G_\Cal A$. \\ D'une part en d\'erivant l'expression ci-dessus par rapport \`a t on a : $ \dot \varphi_{t,\alpha} = \varphi \ast \dot h_{t, \alpha}$\\
D'autre part en d\'erivant la formule \eqref{CHV1} par rapport \`a t on obtient :
\begin{eqnarray*}
\dot \varphi_{t,\alpha} &=& z \alpha \star \varphi_{t,\alpha} \\
&=& \varphi_{t,\alpha} \ast z \wt{\Cal R}_\alpha ( \varphi),
\end{eqnarray*}
d'o\`u: 
\begin{eqnarray*}
\varphi \ast \dot h_{t,\alpha} &=& \varphi_{t,\alpha} \ast z \wt{\Cal R}_\alpha ( \varphi_{t,\alpha})\\
&=& \varphi \ast h_{t, \alpha} \ast z \wt{\Cal R}_\alpha ( \varphi \ast h_{t, \alpha}).
\end{eqnarray*}
$ \wt{\Cal R}_\alpha $ verifie la propri\'et\'e de cocycle \eqref{CHV3}, donc on a :
\begin{eqnarray*}
\varphi \ast \dot h_{t, \alpha} &=& \varphi \ast h_{t, \alpha} \ast z \left( \wt{\Cal R}_\alpha ( h_{t, \alpha}) + h^{\ast_-1} _{t, \alpha} \ast \wt{\Cal R}_\alpha (\varphi )\ast h_{t, \alpha} \right )\\
&=& \varphi \ast h_{t, \alpha} \ast z \wt{\Cal R}_\alpha ( h_{t, \alpha}) + \varphi \ast z \wt{\Cal R}_\alpha (\varphi) \ast h_{t, \alpha}.
\end{eqnarray*}
Donc :
$$ \dot h_{t, \alpha} = h_{t,\alpha} \ast z \wt{\Cal R}_\alpha ( h_{t,\alpha}) +z \wt{\Cal R}_\alpha ( \varphi)\ast h_{t,\alpha}.$$ 
\item Soit $ \varphi \in G^\alpha_\Cal A$. La d\'ecomposition de Birkhoff de $ \varphi_{t,\alpha}$ s'\'ecrit :
\begin{eqnarray*}
\varphi_{t,\alpha} &=& (\varphi_{t,\alpha})^{\ast-1}_- \ast (\varphi_{t,\alpha})_+\\
&=& (\varphi_-)^{\ast-1} \ast (\varphi_{t,\alpha})_+\\
&=& (\varphi \ast \varphi^{\ast-1}_+) \ast (\varphi_{t,\alpha})_+\\
&=& \varphi \ast h_{t, \alpha}.
\end{eqnarray*}
Donc $ h_{t, \alpha} \in G_{\Cal A_+}$. Alors en \'ecrivant la formule \eqref{CHV4} \`a $ t = 0$, on montre que $z \wt{\Cal R}_\alpha ( \varphi)$ appartient \`a $ \mathfrak g^{\Cal A} \cap \Cal L (\Cal H, \Cal A_+)$, ce qui prouve la premi\`ere partie.\\
En \'ecrivant l'\'equation \eqref{CHV4} \`a $ t= 0 $ on obtient :
\begin{equation}
z \wt{\Cal R}_\alpha ( \varphi) =\dot h_\alpha(0) = \frac{d}{dt}_{ | t=0}(\varphi_{t,\alpha})_+ . \label{CHV6}
\end{equation}
Pour $ \varphi \in G^\alpha_{\Cal A_- }= \{ \varphi \in G^\alpha_{\Cal A} \;\; \text{tel que} \;\;  \varphi =\varphi ^{\ast -1}_- \}$ on a, puisque $\varphi ( \mop {Ker} \varepsilon) \subset \Cal A_-$ :
\begin{eqnarray*}
h_{t,\alpha} (x) &=& ( \varphi_{t,\alpha})_+ (x) \\
&=& ( I - \pi ) \left( \varphi_{t,\alpha} ( x ) + \sum_{ x } \varphi^{\ast -1}(x') \varphi_{t,\alpha} ( x'') \right)\\
&=& t ( I - \pi ) \left( z \wt{\Cal R}_\alpha ( \varphi) ( x ) + z \sum_{ x } \varphi^{\ast -1}(x') \wt{\Cal R}_\alpha ( \varphi) ( x'')\right) + O(t^2)\\
&=& t \mop {Res} ( \alpha\star \varphi) + O(t^2).
\end{eqnarray*} 
Donc :
\begin{equation}
\dot h_\alpha (0 ) = \mop {Res} ( \alpha\star \varphi). \label{CHV7}
\end{equation}
Alors d'apr\`es la formule \eqref{CHV6} pour tout $\varphi \in G^\alpha_{\Cal A_-}$ on a : 
\begin{equation}
z \wt{\Cal R}_\alpha ( \varphi) = \mop {Res} ( \alpha \star \varphi). \label{CHV8}
\end{equation}
Inversement, soit $ \chi \in \mathfrak g^c$, on consid\`ere $ \psi = \wt{\Cal R}^{-1}_\alpha ( z^{-1} \chi) $. Cet \'el\'ement de $ G_{\Cal A }$ v\'erifie par d\'efinition, d'apr\`es l'\'equation \eqref{CHV2} :
$$z \psi \circ B_\alpha = \psi \ast \chi.$$
Donc pour tout $ x \in \mop{Ker} \varepsilon $ on a :
\begin{eqnarray*}
\alpha \star \psi (x) &=& \frac{1}{z} \left( \chi (x ) \psi ( 1_\Cal H ) + \psi ( x ) \chi ( 1_\Cal H ) + \sum_{ x } \psi (x') \chi ( x'') \right) \\
&=& \frac{1}{z} \left( \chi (x ) + \sum_{ x } \psi (x') \chi ( x'') \right).
\end{eqnarray*} 
On suppose que $ \alpha \in \mathfrak g^c_\Cal K $ alors $ \alpha \star \psi (x) \in \Cal A_-$. En utilisant la notation de Sweedler on \'ecrit :
\begin{eqnarray*}
\alpha \star \psi (x) &=& \sum_{ x } \alpha (x_0) \psi ( x_1) \\
&=& \alpha ( 1_\Cal K) \psi (x) + \sum_{ x } \alpha (x') \psi ( x'')\\
&=& \psi (x) + \sum_{ x } \alpha (x') \psi ( x'').\\
\end{eqnarray*}
Donc : $$ \psi (x) = \alpha \star \psi (x) - \sum_{ x } \alpha (x') \psi ( x'').$$
Par r\'ecurrence sur le degr\'e de $x$, la formule ci dessus nous permet de montrer que : $$ \psi (x) \in \Cal A_-$$
par suite :
$$ \psi = \wt{\Cal R}^{-1}_\alpha ( z^{-1} \chi) \in G^\alpha_{\Cal A_- }.$$
\item Le deux \'equations $ \varphi_{t,\alpha} = \varphi \ast h_{t, \alpha}$ et $ (\varphi_{t,\alpha})_{s,\alpha} = \varphi_{t+s,\alpha}$ nous permettent d'\'ecrire :
\begin{equation}
h_{s+t, \alpha} = h_{s, \alpha} \ast ( h_{t, \alpha})_{s, \alpha}. \label{CHV9}
\end{equation}
En effet :
\begin{eqnarray*}
h_{s+t, \alpha} &=& \varphi^{\ast-1} \ast (\varphi_{t+s,\alpha})\\
&=& \varphi^{\ast-1} \ast (\varphi_{t,\alpha})_{s,\alpha}\\
&=&\varphi^{\ast-1} \ast \varphi \ast ( h_{t, \alpha} )_{s,\alpha} \\
&=& \varphi^{\ast-1} \ast \varphi_{s,\alpha} \ast ( h_{t, \alpha} )_{s,\alpha}\\
&=& h_{s, \alpha} \ast ( h_{t, \alpha} )_{s,\alpha}.\\
\end{eqnarray*}
En prenant $ z = 0 $ on obtient imm\'ediatement la propri\'et\'e de groupe \`a un param\`etre :
\begin{equation}
F_{s+t , \alpha } = F_{s , \alpha } \ast F_{t , \alpha }. \label{CHV10}
\end{equation}
\end{enumerate}
\end{proof}
Nous pouvons maintenant d\'efinir la foncton $\beta_\alpha$.
\begin{definition}
Pour tout $ \varphi \in G^\alpha_{\Cal A } $ sa fonction $\beta_\alpha $ est le g\'en\'erateur de groupe \`a un param\`etre $ F_{t , \alpha } $, c'est-\'a-dire l'\'element de $\Cal H^\ast$ d\'efini par :
$$\beta_\alpha ( \varphi ) := \frac{d}{dt}_{ | t=0} F_{t , \alpha }(\varphi). $$
\end{definition}
\begin{proposition}
Pour tout $ \varphi \in G^\alpha_{\Cal A } $ sa fonction $\beta_\alpha $ coincide avec celle de la partie n\'egative $ \varphi^{\ast -1}_- $ dans la d\'ecomposition de Birkhoff. Elle est donn\'ee par les expressions :
\begin{eqnarray*}
\beta_\alpha ( \varphi ) &=& \mop {Res} ( \wt{\Cal R}_{\alpha} ( \varphi ))\\
&=& \mop {Res}( \varphi^{\ast -1}_- \circ B_\alpha ) \\
&=& - \mop {Res}( \varphi_- \circ B_\alpha ). 
\end{eqnarray*}
\end{proposition}
\begin{proof}[Preuve]
On suppose que $ \varphi \in G^\alpha_{\Cal A_- } $ , alors $ \varphi^{\ast -1}_- = \varphi $. D'o\`u d'apr\`es la proposition \ref{transposeL}, $ z \wt{\Cal R}_{\alpha} ( \varphi )$ est constant. La proposition r\'esulte alors des \'equations : $$ \dot h (0 ) = \mop {Res} ( \varphi \circ B_\alpha ) \;\;\text{ et } \;\; z \wt{\Cal R}_{\alpha} ( \varphi ) = \mop {Res} ( \varphi \circ B_\alpha ).$$
On suppose que $ \varphi \in G^\alpha_{\Cal A } $ et considerons la d\'ecomposition de Birkhoff. Comme $ \varphi^{\ast -1} $ et $ \varphi_{+} \in G^\alpha_{\Cal A } $, en appliquant la proposition \ref{transposeL} nous obtenons :
$$
\varphi_{ t , \alpha } = \varphi \ast h_{ t , \alpha } $$
$$ ( \varphi^{\ast{-1}}_- )_{ t , \alpha } = \varphi^{\ast {-1}}_- \ast v_{ t , \alpha } $$
$$ ( \varphi_+ )_{ t , \alpha } = \varphi_+ \ast w_{ t , \alpha }, $$
et l'egalit\'e : $$ \varphi_{ t , \alpha } = ( \varphi^{\ast{-1}}_- )_{ t , \alpha } \ast ( \varphi_+ )_{ t , \alpha } $$
donne :
\begin{eqnarray*}
h_{ t , X } &=& \varphi^{\ast{-1}} \ast \varphi _{ t , \alpha }\\
&=& \varphi^{\ast{-1}} \ast ( \varphi^{\ast{-1}}_- )_{ t , \alpha } \ast ( \varphi_+)_{ t , \alpha } \\
&=& \varphi^{\ast{-1}} \ast ( \varphi^{\ast{-1}}_- ) \ast v_{ t , \alpha} \ast \varphi_+ \ast w_{ t , \alpha}.
\end{eqnarray*}
Donc :
\begin{equation}
h_{t, \alpha} = ( \varphi _+)^{\ast{-1}} \ast v_{ t , \alpha} \ast \varphi _+ \ast w_{t , \alpha}. \label{CHV11}
\end{equation}
On d\'esigne par $ F_{t , \alpha } , V_{t , \alpha } , W_{t , \alpha } $ resp\'ectivement les sous-groupes \`a un param\`etre obtenus par $ h_{t , \alpha } , v_{t , \alpha} , w_{t , \alpha } $ \`a $ z = 0 $, il est clair que $( \varphi_{+} ) _ {| z =0 } = e $ et de m\^eme $W_{t , \alpha } $ est un groupe \`a un param\`etre constant reduit \`a l'\'el\'ement neutre $ e$. Ainsi l'\'equation \eqref{CHV11} \`a $ z = 0$ se r\'eduit en : 
$$ F_{t , \alpha } = V_{t , \alpha }. $$
Ce qui prouve la pr\'emi\`ere affirmation. En utilisant la propri\'et\'e de cocycle on obtient :
$$ \wt{\Cal R}_\alpha (\varphi ) = \wt{\Cal R}_\alpha (\varphi^{\ast -1}_- \ast \varphi_+) = \wt{\Cal R}_\alpha ( \varphi_+) + \varphi^{\ast -1}_+ \ast \wt{\Cal R}_\alpha (\varphi^{\ast -1}_-)\ast \varphi_+ .$$
Donc :
$$ \mop {Res}\left( \wt{\Cal R}_\alpha (\varphi )\right ) =\mop {Res} \left( \wt{\Cal R}_\alpha ( \varphi^{\ast -1}_- )\right).$$
Comme :
$$ \wt{\Cal R}_\alpha( e) = \wt{\Cal R}_\alpha ( \varphi_- \ast \varphi^{\ast -1}_-) = \wt{\Cal R}_\alpha ( \varphi_-) + \varphi^{\ast -1}_- \ast \wt{\Cal R}_\alpha (\varphi^{\ast -1}_-)\ast \varphi_- = 0. $$
Donc : $$ \wt{\Cal R}_\alpha (\varphi^{\ast -1}_-) = - \varphi^{\ast -1}_- \ast \wt{\Cal R}_\alpha (\varphi_-)\ast \varphi_- .$$
D'o\`u : $$ \mop {Res}( \varphi^{\ast -1}_- \ast B_\alpha ) = - \mop {Res}( \varphi_- \ast B_\alpha).$$
\end{proof}

\end{document}